# Regulating nanoscale heat transfer with Janus nanoparticles

*Chen Xie,[†] Blake Wilson,[†] Zhenpeng Qin\*,[†,§,∥]*

[†]Department of Mechanical Engineering, University of Texas at Dallas, 800 West Campbell Road, Richardson, Texas 75080, United States.

[§]Department of Bioengineering, Department of Bioengineering, Center for Advanced Pain Studies, University of Texas at Dallas, 800 West Campbell Road, Richardson, Texas 75080, United States.

[∥]Department of Surgery, University of Texas at Southwestern Medical Center, 5323 Harry Hines Boulevard, Dallas, Texas 75390, United States.

**Abstract**

Janus nanoparticles (JNPs) with heterogeneous compositions or interfacial properties can exhibit directional heating upon external excitation, such as laser radiation and magnetic field. This directional heating may be harnessed for new nanotechnology and biomedical applications. Understanding thermal transport and temperature control with JNP heating is critical for these advances. Here, we developed a numerical framework to analyze the asymmetric thermal transport in JNP heating under photothermal stimulation. We found that JNP-induced temperature contrast, defined as the ratio of temperature increase in the surrounding water, shows a substantial size and polar angle dependence. Notably, we discovered a significant enhancement of the temperature contrast under pulsed heating due to thermal confinement, compared with the continuous heating. This work brings new insights into the thermal responses of JNP heating and advances the field.

**Main context:**

**Introduction:**

In the past decade, remarkable advances have been made with the development of nanoparticle heating,[1-4] including cancer treatment,[5-9] neuron modulation,[10, 11] molecular hyperthermia,[12-14] point-of-care diagnostic,[15] imaging enhancement,[16-18] energy harvesting,[19, 20] and catalysis.[21-23] Janus nanoparticles (JNPs) have recently brought new opportunities to the field.[24, 25] JNPs are novel nanostructures consisting of heterogeneous structures and materials, including combinations of metallic and non-metallic materials.[26-28] The asymmetric coating on JNPs can lead to heterogeneous interfacial properties, as an example, hydrophobic versus hydrophilic with high versus low interfacial thermal resistances (ITRs), respectively.[29] This heterogeneity of interfacial properties can be used to regulate the thermal transport around the JNP and lead to directional heating.[30, 31] The directional heating has enabled novel applications with JNPs, such as thermophoretic nanomotor,[31-33] nano tweezer,[34, 35] Janus nano pen injection,[36] and energy management.[37]

The central issue for JNP heating is a better understanding of thermal transport.[35] Olarte-Plata et al. analyzed the directional heating of JNPs with heterogeneous coating and ITRs using molecular dynamic simulation and demonstrated a temperature difference (~10K) with octanethiolate/mercaptohexanol-coated JNPs.[30] Besides this modeling work, Xuan et al. have developed a self-driven thermophoretic nanomotor with continuous JNP heating. They have shown the speed of the nanomotor can reach ~40 μm/s with a laser intensity of 60 W/cm².[33] While these studies have greatly advanced our understanding of directional heating with JNP and demonstrated promising applications, it remains unclear how heterogeneous interface properties of JNP and laser excitation (continuous vs pulsed) contribute to regulating the nanoscale thermal transport.

In this work, we developed a finite difference model to elucidate how the heterogeneous interface properties and laser excitation method regulate the nanoscale heat transfer (Fig 1A). we defined a dimensionless parameter based on the ratio of asymmetric temperature increase ($\zeta = \frac{\Delta T_1}{\Delta T_2}$, Fig 1B, where $\Delta T_1$ and $\Delta T_2$ indicates the representative temperature rise of the two coating parts.) to quantify the directional heating. Our results suggest that increasing JNP size reduces the temperature contrast. We found that the distribution of interface thermal resistance led to an asymmetric impact on the temperature contrast. Importantly, we demonstrate a significant enhancement of the temperature contrast with pulsed JNP heating compared with continuous heating. Our work advances the understanding of directional heating with JNP and brings new perspectives and potential applications to the field.

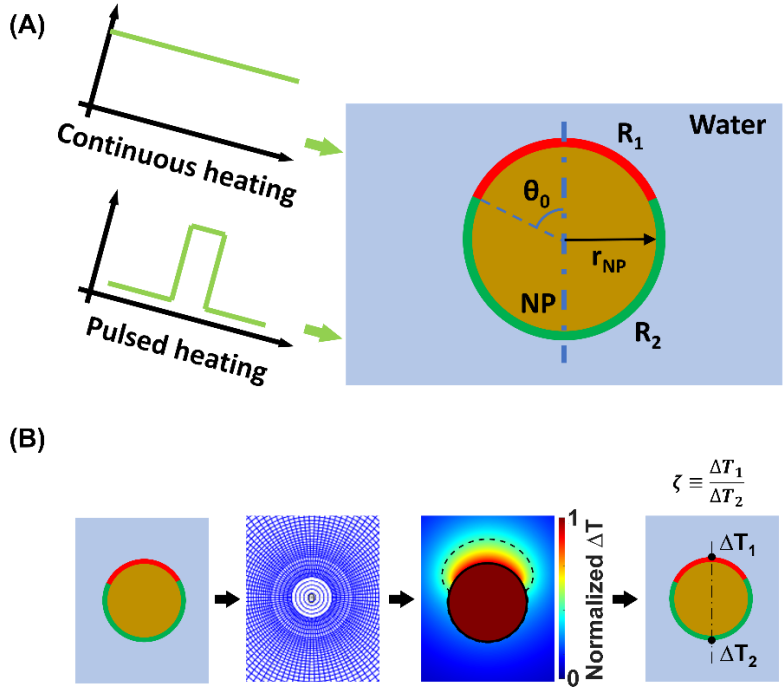

**Figure 1. Schematic illustration of the modeling of JNP heating.** (A) A spherical Janus nanoparticle (JNP) is heterogeneously coated, leading to nonuniform interfacial thermal resistance (ITR). The two ITRs on the JNP are $R_1$, covering the north pole, and $R_2$, covering the south pole. The polar angle ($\theta_0$) is defined as the critical cone angle that pinpoints the boundary between two ITRs. We impose continuous and pulsed heating on the JNP to model the steady-state and transient heating scenarios. (B) We developed an FDM numerical model and discretized the system under spherical coordinates. We defined the dimensionless parameter ($\zeta = \frac{\Delta T_1}{\Delta T_2}$) to characterize the directional heating.

**Method:**

**1  Theoretical model:**

We start with establishing the theoretical model. Here we consider a plasmonic JNP with a gold core, asymmetric coating on its surface, and thus heterogeneous ITR (Fig. 1A) along the interface. The JNP is submerged in an aqueous solution, and heat can transport through the interface into the water when the JNP is excited by laser radiation.[35] Recently, studies reported that heat conduction under the nanoscale could deviate from Fourier's law when the mean free path is comparable to the characteristic length.[38] In this work, the temperature profile in the water domain is our primary concern. Considering that the mean free path in the water (~0.1nm) is orders of magnitude smaller than the characteristic length (~10nm), the Fourier law should be sufficient to model the thermal transport[14, 39], We applied Fourier's law with the governing equation[40]:

$$\nabla \cdot (k\nabla T) + g = \rho c \frac{\partial T}{\partial t} \qquad (1)$$

Here, $k$ is the thermal conductivity, $g$ is the heat source, $\rho$ is density, and $c$ is the specific heat. We further expand the governing equation under spherical coordinates with azimuthal symmetry and simplify our model to a 2D problem (Fig. 1A & Fig S1). To further simplify the model, we assume constant thermal properties. Based on our previous work, the constant property assumption introduces minimal error.[14] The theoretical model is shown as follows:

$$\frac{\partial^2 T}{\partial r^2} + \frac{2}{r}\frac{\partial T}{\partial r} + \frac{1}{r}\frac{\partial^2 T}{\partial \theta^2} + \frac{cot(\theta)}{r^2}\frac{\partial T}{\partial \theta} + \frac{g}{k} = \frac{1}{D}\frac{\partial T}{\partial t} \qquad (2)$$

where $D$ refers to thermal diffusivity. Expanding the governing equation with spherical coordinates enables us to discretize the system with an orthogonal mesh (Fig. 1B & Fig S1). However, it also has some drawbacks. As shown in Equation (2), the system will have singularities when $r \to 0$ and $\theta \to 0 \ \& \ \pi$ and is discontinuous at the center point where the two singularities meet ($r = 0, \theta = 0$ ).[41] To overcome this, we assume the temperature is uniformly distributed in the $\theta$ direction near the center point. Thus we can further reduce the governing equation in the region near the center point to a 1D problem (Fig S2). Since the thermal diffusivity of metal is much higher than that in water, the temperature profile inside the JNP is close to isothermal, and this 1D assumption is acceptable. Consequently, the temperature governing equation is given by:

$$\begin{cases} \dfrac{\partial^2 T_{JNP}}{\partial r^2} + \dfrac{2}{r}\dfrac{\partial T_{JNP}}{\partial r} + \dfrac{g}{k_{JNP}} = \dfrac{1}{D_{JNP}}\dfrac{\partial T_{JNP}}{\partial t}, & r \leq r_{1D} \\ \dfrac{\partial^2 T_{JNP}}{\partial r^2} + \dfrac{2}{r}\dfrac{\partial T_{JNP}}{\partial r} + \dfrac{\cot(\theta)}{r^2}\dfrac{\partial T_{JNP}}{\partial \theta} + \dfrac{1}{r}\dfrac{\partial^2 T_{JNP}}{\partial \theta^2} + \dfrac{g}{k_{JNP}} = \dfrac{1}{D_{JNP}}\dfrac{\partial T_{JNP}}{\partial t}, & r_{1D} < r \leq r_{JNP} \\ \dfrac{\partial^2 T_w}{\partial r^2} + \dfrac{2}{r}\dfrac{\partial T_w}{\partial r} + \dfrac{\cot(\theta)}{r^2}\dfrac{\partial T_w}{\partial \theta} + \dfrac{1}{r}\dfrac{\partial^2 T_w}{\partial \theta^2} = \dfrac{1}{D_w}\dfrac{\partial T_w}{\partial t}, & r > r_{JNP} \end{cases} \quad (3)$$

The boundary and initial conductions are:

$$\begin{aligned} \dfrac{T_{JNP} - T_w}{R} &= -k_{JNP}\dfrac{\partial T_{JNP}}{\partial r} = -k_w\dfrac{\partial T_w}{\partial r}, \quad r = r_{JNP} \\ T_w &= 0, \quad r = r_{boundary} \\ T_{JNP} &= 0, T_w = 0, t = 0 \end{aligned} \quad (4)$$

$r_{1D}$ refers to the radius of the 1D region (Fig S2), $r_{JNP}$ refers to the radius of the JNP, and $r_{boundary}$ refers to the radius of the boundary for the water domain. It should be noted that $R$ is the interfacial thermal resistance (ITR), and it is a function of $\theta$:

$$R = \begin{cases} R_1, & 0 \leq \theta \leq \theta_0 \\ R_2, & \theta_0 < \theta \leq \pi \end{cases} \quad (5)$$

where $R_1$ and $R_2$ indicate the ITR for the coating on the northern part and the ITR for the coating on the southern part, respectively, and $\theta_0$ refers to the polar angle boundary between the heterogenous coatings (Fig. 1A).

Equation (3)-(5) are the mathematical description of our model, and we further developed our FDM model based on Equation (3)-(5).

## 2    Finite difference method model

Next, we developed our (Finite difference method) FDM model by discretizing Equation (3)-(5).[42] Here, we adopted a biased mesh in the radius direction in the water domain to achieve higher accuracy and save computational time (Fig. S1). For the steady-state problem, the left-hand side of Equation (3) equals zero, and the discretized governing equation in the water domain is as follows:

$$\frac{A_a T_{i-1,j} + A_b T_{i+1,j} + A_c T_{i+2,j} + (A_a + A_b + A_c)T_{i,j}}{(\Delta r)^2}$$
$$+ \frac{2}{r_i}\frac{D_a T_{i+1,j} + D_b T_{i-1,j} + (D_a + D_b)T_{i,j}}{\Delta r} + \frac{1}{r_i}\frac{T_{i,j+1} - 2T_{i,j} + T_{i,j-1}}{(\Delta\theta)^2}$$
$$+ \frac{\cot(\theta_j)}{(r_i)^2}\frac{T_{i,j+1} - T_{i,j-1}}{2\Delta\theta} = 0 \quad (6)$$

$$A_a = \frac{2(\alpha^3 + 2\alpha^2 + \alpha - 1)}{\alpha^4 + 3\alpha^3 + 2\alpha^2 + \alpha + 1}, A_b = \frac{2(\alpha^2 + \alpha - 1)}{\alpha^4 + 2\alpha^3 - \alpha^2}, A_c = \frac{-2(\alpha - 1)}{\alpha^7 + 4\alpha^6 + 5\alpha^5 + 3\alpha^4 - \alpha^2}$$
$$D_a = \frac{-\alpha}{\alpha + 1}, D_b = \frac{1}{\alpha^2 + \alpha}$$

where the subscript $i$ indicates the $i$th node in the $r$ direction, and $j$ indicates the $j$th node in $\theta$ direction, $\alpha$ indicates the biased mesh factor ($\Delta r_{i+1} = \alpha \Delta r_i$). The system has a second order of accuracy, and we validated the model by conducting a mesh-independent analysis and boundary effect analysis (Fig. S2 & S3).

For the transient problem, we discretized Equations (3)-(5) with a Crank-Nicolson scheme,[43] and the discretized governing equation in the water domain is:

$$\frac{1}{2}\left[\frac{A_a T_{i,j}^{n+1} + A_b T_{i+1,j}^{n+1} + A_c T_{i+2,j}^{n+1} + (A_a + A_b + A_c)T_{i,j}^{n+1}}{(\Delta r)^2}\right.$$
$$\left.+\frac{A_a T_{i,j}^n + A_b T_{i+1,j}^n + A_c T_{i+2,j}^n + (A_a + A_b + A_c)T_{i,j}^n}{(\Delta r)^2}\right]$$
$$+\frac{1}{r_i}\left[\frac{D_a T_{i+1,j}^{n+1} + D_b T_{i-1,j}^{n+1} + (D_a + D_b)T_{i,j}^{n+1}}{\Delta r}\right.$$
$$+\frac{D_a T_{i+1,j}^n + D_b T_{i-1,j}^n + (D_a + D_b)T_{i,j}^n}{\Delta r}\right]$$
$$+\frac{1}{r_i}\left[\frac{T_{i,j+1}^{n+1} - 2T_{i,j}^{n+1} + T_{i,j-1}^{n+1}}{(\Delta\theta)^2} + \frac{T_{i,j+1}^n - 2T_{i,j}^n + T_{i,j-1}^n}{(\Delta\theta)^2}\right]$$
$$+\frac{\cot(\theta_j)}{2(r_i)^2}\left[\frac{T_{i,j+1}^{n+1} - T_{i,j-1}^{n+1}}{2\Delta\theta} + \frac{T_{i,j+1}^n - T_{i,j-1}^n}{2\Delta\theta}\right] = \frac{1}{D_w}\frac{T_{i,j}^{n+1} - T_{i,j}^n}{\Delta t} \quad (7)$$

where superscript *n* indicates the *n*th node in the time direction, we further validated the accuracy and stability of the model (Fig. S4), the numerical results are in consistent with analytical solutions. Equation (6) and (7) are solved with MATLAB R2020b.

**Results and discussion**

**1      Directional heating and temperature contrast under continuous heating (steady-state)**

Several factors can affect the thermal transport and temperature profile during JNP heating, including the interfacial thermal resistance (ITR), the size of the JNP ($r_{JNP}$), the polar angle of the heterogeneous coatings ($\theta_0$), and the heating time (*t*). Here in this section, as we focus on the directional heating under steady-state scenarios, we start with the most idealized JNP with a $\theta_0 = \pi/2$ (Fig. 1A), i.e., the JNP is heterogeneously coated equally on the two hemispheres. Nonuniform interfacial properties can stem from the heterogeneous coating, including hydrophilicity and ITR, and the contrast between ITRs can be tuned by altering the combination of coating materials or ligands.[44-46] Here, we analyzed the temperature profile with different combinations of the ITRs, where $R_1$ and $R_2$ indicate the ITRs along the northern and southern hemispheres, respectively. Fig. 2A shows that for a given sized JNP ($r_{JNP}$ = 15 nm), the combination of $R_1$ and $R_2$ significantly affects the temperature profile. For case ① where $R_1 = R_2 = 50 \times 10^{-9}$ $m^2KW^{-1}$, we have a homogeneous nanoparticle (NP) with uniform ITR; thus, the temperature is independent of the polar angle ($\theta$), and a uniform temperature jump is observed across the JNP-water interface due to ITR. As we increase the contrast between ITRs, a temperature contrast in the water domain near the north pole and south pole gradually appears, and a strong positive relationship between the ITR contrast and temperature contrast is illustrated in Fig. 2A, an approximately two times difference between the temperature around the north pole and the south pole is observed for case ④ ($R_2/R_1 = 100$). This heterogeneous thermal transport and temperature profile are known as directional heating. Here, to quantify the temperature contrast and directional heating, we developed a dimensionless parameter $\zeta$ which is defined as the ratio between the temperature rise in water at the north pole ($\Delta T_1$) and the south pole ($\Delta T_2$) (Fig. 1B):

$$\zeta = \frac{\Delta T_1}{\Delta T_2} \qquad (8)$$

As shown in Equation (8), $\zeta$ quantifies the temperature contrast for directional heating by characterizing the relative difference between the temperature rises. For $\zeta = 1$, it indicates no temperature contrast or homogeneous heating, as shown in Fig. 1A, case ①. On the contrary, when $\zeta$ deviates from 1, it indicates considerable temperature contrast and directional heating (Fig. 1A, case ④ $\zeta = 2.5$).

For more generalized cases with arbitrary combinations of $R_1$ and $R_2$, we demonstrated and investigated the 2D map of $\zeta_{s.s.}$ in terms of $R_1$ and $R_2$ (Fig. 2B), where the subscript s.s. indicating it's under steady-state. The range for ITRs in Fig. 2B (1-100 × $10^{-9}$ $m^2KW^{-1}$) is consistent with ITRs reported in the literatures.[14] $\zeta_{s.s.}$ equals to 1 along the diagonal where $R_1 = R_2$. On the contrary, $\zeta_{s.s.}$ deviates from 1 in the off-diagonal region, denoting directional heating when $R_1 \neq R_2$. It should be noted that Fig. 2B is symmetry along the diagonal line due to $\theta_0 = \pi/2$ (Fig. 1A), i.e., for an idealized JNP where the two coating areas are equal, exchange the $R_1$ and $R_2$ should lead to the inverse of $\zeta_{s.s.}$. On this $\zeta_{s.s.}$-ITR map, we can readily read the relative temperature contrast with arbitrary combinations of $R_1$ and $R_2$; for instance, when $R_1 = 1 \times 10^{-9}$ m$^2$KW$^{-1}$ and $R_2 = 100 \times 10^{-9}$ m$^2$KW$^{-1}$, $\zeta_{s.s.} = 2.5$, indicating significant directional heating, with the temperature rise near the north pole being 2.5 times higher than that near the south pole.

To further investigate how the combination of ITRs affects $\zeta_{s.s.}$, we developed a circuit analogy model to analyze the JNP heating under steady-state analytically. Here we assume no heat transfer across the polar angle ($\theta_0$) of heterogeneous ITR in the water domain, and we simplified the model to a parallel circuit analogy (Fig. S5A) with a set of linear equations to describe the temperature at the critical nodes (Equation S1 & S2). Based on this parallel circuit model, $\zeta_{s.s.}$ can be estimated by:

$$\zeta_{s.s.} = \frac{\Delta T_1}{\Delta T_2} = \frac{(r_{JNP} + R_2 k_w)}{(r_{JNP} + R_1 k_w)} \qquad (9)$$

Comparing results from Equation (9) with the FDM results (Fig. S5B), we show that the parallel circuit model overestimates $\zeta_{s.s.}$, but is more accurate with larger JNPs. This overestimation may have arisen from the assumption of no heat transfer across $\theta_0$, and for larger JNPs, heat flux on the $\theta$ direction becomes less important, making it more accurate for larger JNPs. Overall, Equation (9) helps us better understand JNP heating and demonstrates the size dependence of temperature contrast with JNP heating. Based on Equation (9), for a given combination of $R_1$ and $R_2$, increasing $r_{JNP}$ will yield less temperature contrast. This size dependence agrees with Kapitza number analysis,[47, 48] where for larger NP, the ITR has an insignificant effect on the heat dissipation. This size dependence of $\zeta_{s.s.}$ is also confirmed with the FDM model, and the $\zeta_{s.s}$ maps with $r_{JNP} =$ 3.75 nm and $r_{JNP} =$ 60 nm are shown in Fig. S6.

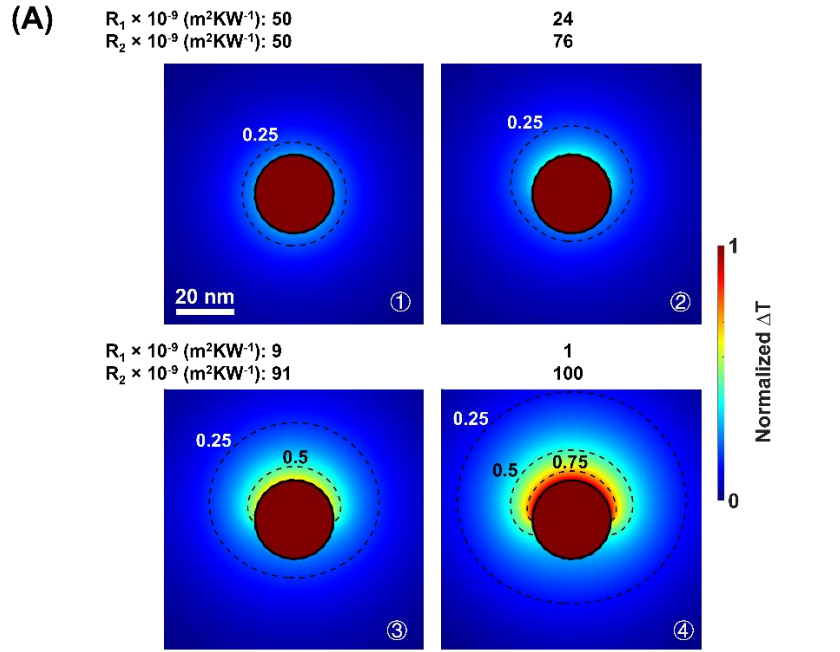

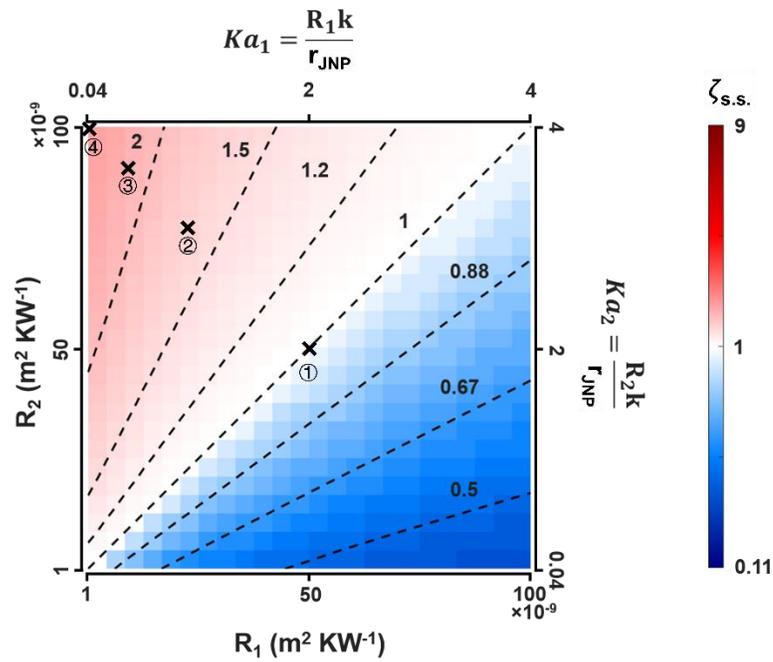

**Figure 2. Temperature profile and temperature contrast with JANUS NP under S.S.**
(A) Profiles for normalized temperature rise ($\Delta T$) with Janus nanoparticle (JNP) heating. The JNP size ($r_{JNP}$) is 15 nm, and the cone angle for the heterogeneous coating ($\theta_0$) is $\pi/2$. (B) 2D map of $\zeta_{s.s.}$ in terms of the two ITRs, $R_1$, and $R_2$. Dotted lines are contour lines for $\zeta_{s.s.}$.

## 2 The effect of JNP size and polar angle on the temperature contrast under S.S.

Next, we focused on the effect of polar angle ($\theta_0$) on temperature contrast under steady-state heating. In previous sections, we discussed the directional heating and temperature contrast with idealized JNPs ($\theta_0 = \pi/2$). In reality, $\theta_0$ can deviate from $\pi/2$ depending on the synthesis protocol (Fig. S7).[28, 49] Here, we found that $\theta_0$ has a significant impact on the temperature profile as well as on $\zeta_{s.s.}$. As illustrated in Fig. S8, for a 30 nm JNP with $R_2/R_1 = 100$, when $\theta_0 = 0$, the JNP is homogeneously coated with ITR of $R_2$, leading to $\zeta_{s.s.} = 0$ (homogeneous heating). As we slightly increase $\theta_0$, $\zeta_{s.s.}$ rises drastically and reaches its peak value (~4.2) at $\theta_0 \sim \pi/12$. After that peak point, $\zeta_{s.s.}$ gradually drops as we further increase $\theta_0$, and eventually down to 1 when $\theta_0 = \pi$, indicating homogenous heating as uniform ITR = $R_1$. Here we could observe an asymmetric effect of the $\theta_0$ on $\zeta_{s.s.}$ as the maximum value of $\zeta_{s.s.}$ is achieved for $\theta_0 < \pi/2$ and $\zeta_{s.s.}(\theta_0 = \pi/4) \neq \zeta_{s.s.}(\theta_0 = 3\pi/4)$. The asymmetric effect of $\theta_0$ on $\zeta_{s.s.}$ is also demonstrated in Fig. S8.

This asymmetric effect of $\theta_0$ on $\zeta_{s.s.}$ can be explained by the heterogeneity of local heat flux rate ($\dot{q}$) across the interface. The total heat flux ($q$) across the JNP-water interface only depends on the heating power within JNP for steady-state scenarios and can be treated as a constant. On the contrary, the local heat flux rate ($\dot{q}$) across the JNP-water interface is heavily impacted by the heterogeneous ITR. For JNPs with a high contrast of ITRs, a major fraction of heat flux is through the interface with low ITR (Fig. S9B), i.e., heat flux is regulated through a window with low ITR. As we change $\theta_0$, that window with low resistance can be narrowed or broadened, hence changing the local thermal flux rate (Fig. S8B), and a much greater local heat flux rate at the north pole ($\dot{q}_1$) is observed for smaller $\theta_0$ whereas the local heat flux rate at the south pole ($\dot{q}_2$) is less sensitive to $\theta_0$. Consequently, greater temperature contrast comes with smaller $\theta_0$ as the local temperature gradient is linearly proportional to the local heat flux rate.

It should be noted that for different-sized JNPs and JNPs with less ITR contrasts, all the key factors, including the ITR contrast, the JNP size, and $\theta_0$, can have a combined nonlinear effect on the temperature contrast. Here we demonstrate this combined effect by plotting the map of $\zeta_{s.s.}$ in terms of $r_{JNP}$ and $\theta_0$ with different ITR contrast under the steady state (Fig. 3). Fig. 3 demonstrates that the temperature contrast is more sensitive to $r_{JNP}$ and $\theta_0$ with high ITR contrast and vice versa. Within the range of our parameter scanning, the greatest temperature contrast ($\zeta_{s.s.} = 10$) under steady-state is observed for a 3 nm JNP with $\theta_0 \sim \pi/12$ and a high ITR contrast ($R_2/R_1 = 100$). Our results provide estimations of the directional heating under realistic physical constraints, and we expect Fig. 3 can provide guidelines for tuning the temperature contrast with JNP heating for novel applications.

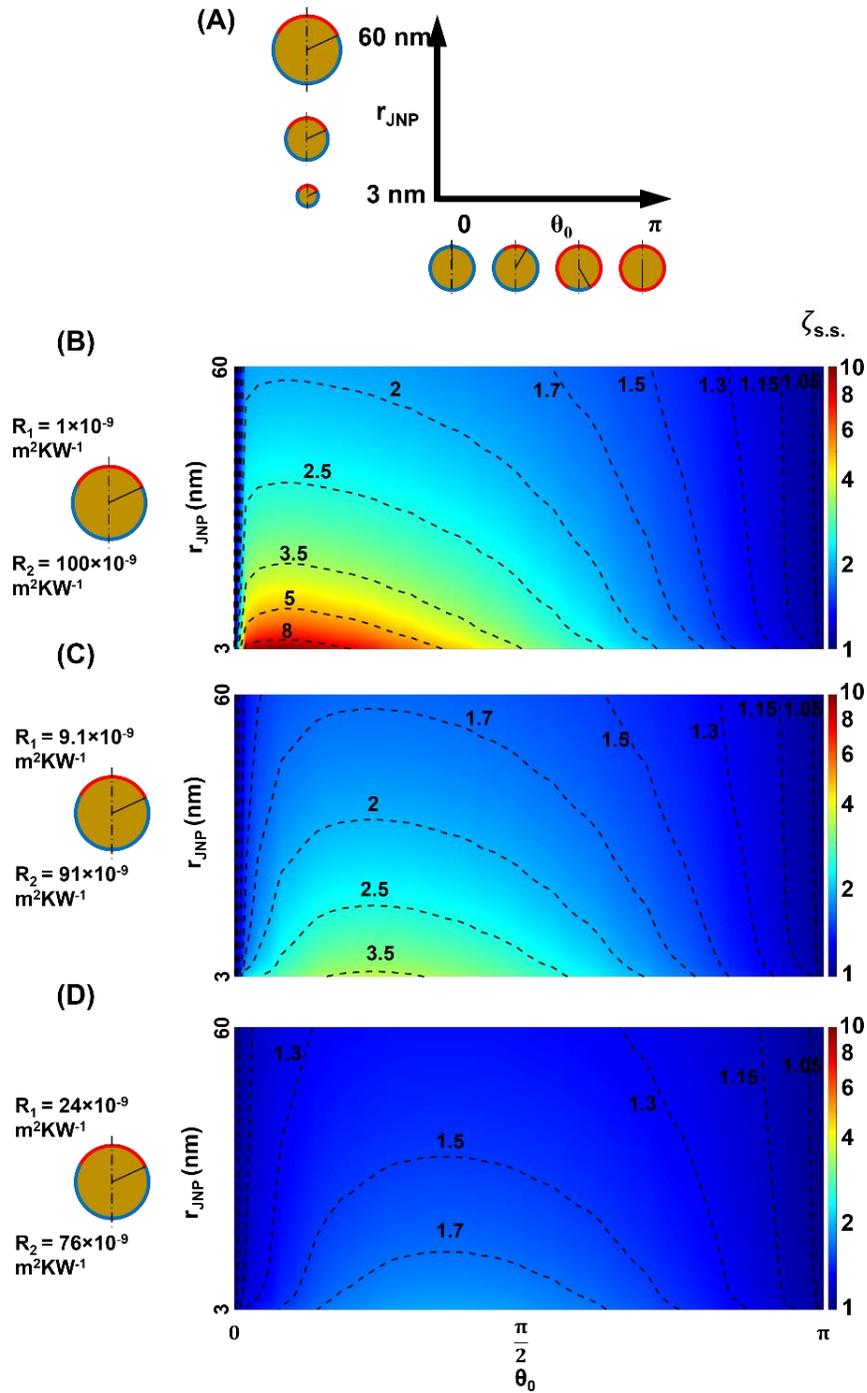

**Figure 3. $\zeta_{s.s.}$ map in terms of $r_{JNP}$ and $\theta_0$ with different ITR contrast under steady-state.** (A) Schematic illustration of altering the $r_{JNP}$ and $\theta_0$. (B-D) $\zeta_{s.s.}$ map in terms of $r_{JNP}$ and $\theta_0$ with (B) $R_1 = 1\times10^{-9}$ m$^2$KW$^{-1}$, $R_2 = 100\times10^{-9}$ m$^2$KW$^{-1}$, (C) $R_1 = 9.09\times10^{-9}$ m$^2$KW$^{-}$

$^1$, $R_2 = 90.9 \times 10^{-9}$ m$^2$KW$^{-1}$, (D) $R_1 = 9.09 \times 10^{-9}$ m$^2$KW$^{-1}$, $R_2 = 90.9 \times 10^{-9}$ m$^2$KW$^{-1}$. The dotted line are contour lines for $\zeta_{s.s.}$.

## 3 Directional heating during pulsed heating

In previous sections, we have discussed the directional heating and temperature contrast with JNP heating under the steady state, which is applicable for a microsecond or continuous heating. Recently, ultra-fast NP heating with femtosecond to nanosecond pulsed energy excitation has shown unique thermal responses and has played essential roles in various applications.[50-52] As the characteristic thermal relaxation time ($\tau = \frac{r_{JNP}^2}{D_w}$) for 30 nm JNP is around 1.6 ns, JNP heating with nanosecond or shorter pulsed heating is far from steady-state. Therefore, in this section, we analyzed the evolution of temperature contrast with JNP heating during the transient process.

Here we Imposed 100 ns pulsed constant heating on a 30 nm JNP with $R_2/R_1 = 100$ and $\theta_0 = \pi/2$. Fig. 4A illustrates the temperature profiles during the transient heating process. At t = 1 ns, due to the short heating time, only a thin layer of water adjacent to the JNP-water interface is heated along the northern hemisphere, also known as thermal confinement. On the contrary, no apparent heating is observed near the southern hemisphere due to greater ITR. By comparing the local heat flux at the north pole and south pole ($\dot{q}_1$ and $\dot{q}_2$), we can see that both the magnitude and increasing rate for $\dot{q}_1$ are significantly higher than those for $\dot{q}_2$ for $t < 1$ ns, this leads to a much-enhanced relative difference between $\Delta T_1$ and $\Delta T_2$ (Fig. 4B), i.e., enhanced temperature contrast. Fig. 4C shows that $\zeta$ can reach to ~20 when $t = 0.1$ ns. As heating continuous, thermal energy continuously diffuses, and both $\Delta T_1$ and $\Delta T_2$ rise; However, the increasing rate for $\Delta T_2$ is greater than that of $\Delta T_1$, ending up with a reducing $\zeta$ as demonstrated in Fig. 4C. For $t =$ 100 ns, the maximum temperature is reached as it is the end time point for the pulsed heating (Fig. 4A). At this time-point, $\Delta T_1$ and $\Delta T_2$ asymptotically tend to its steady-state value (Fig. 4B, $Fo = 63.5$), and $\zeta$ asymptotically falls to its steady-state level (Fig. 4C, $\zeta_{s.s.}$ = 2.5). After the pulsed heating ($t > 100$ ns), the hole system cools down rapidly, and both $\Delta T_1$ and $\Delta T_2$ drop to 0 with $\zeta$ falls to 1. To summarize, we found that $\Delta T_2$ is minimal for small $Fo$, leading to a significant enhancement of temperature contrast, and eight times enhancement of $\zeta$ is observed when $Fo = 0.06$ ($R_2/R_1 =100$, $r_{JNP} = 15$ nm, $\theta_0 = \pi/2$), i.e., pulsed JNP heating can lead to a greater temperature contrast.

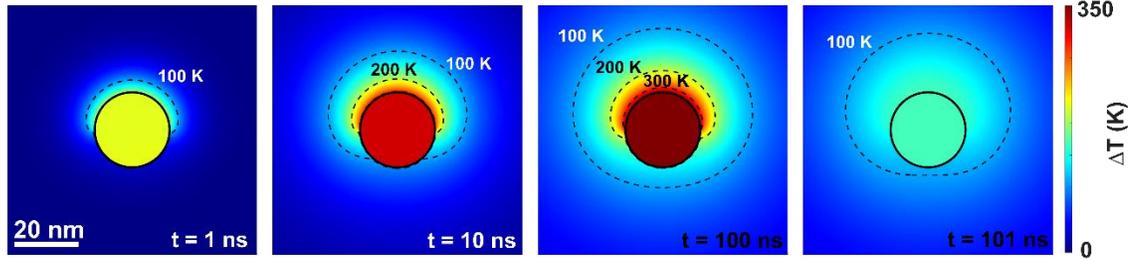

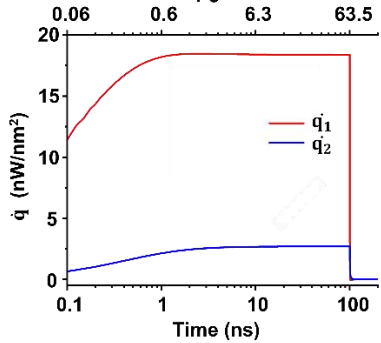 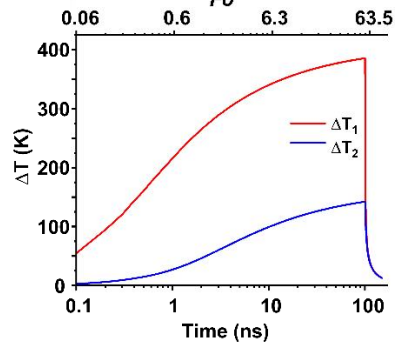 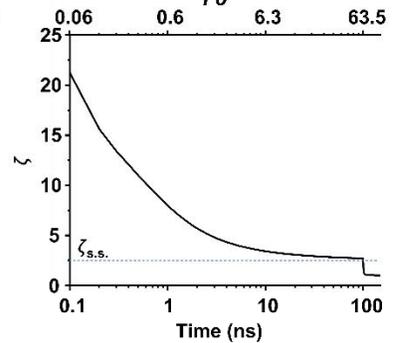

**Fig. 4 The temperature evolution during the transient heating.** (A) $\Delta T$ profiles during the transient heating process with JNP. $r_{JNP} = 15$ nm, heating power ($g$) =35.6 µW,

heating time ($t_{heating}$) = 100 ns. $R_1 = 1 \times 10^{-9}$ m$^2$KW$^{-1}$, $R_2 = 100 \times 10^{-9}$ m$^2$KW$^{-1}$, $\theta_0 = \pi/2$.

(B) The evolution of local heat flux rate ($\dot{q}$) at the north pole and south pole during the transient heating process. (C) The evolution of $\Delta T_1$ and $\Delta T_2$ during the transient heating process. (D) The evolution of $\zeta$ during the transient heating process, the blue dotted line indicates the corresponding $\zeta_{s.s.}$.

## 4 The effect of $r_{JNP}$ and $\theta_0$ on temperature contrast during the transient process

In the previous section, we discussed directional heating during the transient heating process and found an enhancement of temperature contrast with puled JNP heating. This section will expand our investigation into more generalized cases and discuss the combined effects of geometric parameters ($r_{JNP}$ and $\theta_0$) on temperature contrast during the transient heating process.

We start with the effect of the size of the JNP ($r_{JNP}$) on $\zeta$ during the transient process. In our previous sections, we demonstrated a strong size dependence of $\zeta$ for the steady-state conditions (Fig. 3-D and Fig. S6). As shown in Fig. 5A, we can observe a strong size dependence for $\zeta$ throughout the transient process as well. Smaller JNPs are more sensitive to heating time and hence have a stronger enhanced temperature contrast with pulsed heating. Additionally, we analyzed the effect of $\theta_0$ in the transient process. Fig. 5B illustrates the 2D map of $\zeta$ in terms of $t$ and $\theta_0$. For $\theta_0$ close to 0 and $\pi$, JNP turns into homogeneous NP, and $\zeta$ keeps close to 1 throughout the transient heating process. For a moderate $\theta_0$ ($\pi/12 < \theta_0 < 3\pi/4$), we observe an enhancement of $\zeta$ for shorter heating time. Unlike $r_{JNP}$, which has a monotonic effect on $\zeta$ throughout time, $\zeta$ is less sensitive to $\theta_0$ at a low $Fo$ number (~0.1) within this medium $\theta_0$ range. As heating continues, $\zeta$ becomes more sensitive to $\theta_0$ as it tends to the steady-state scenario shown in Fig S8. Lastly, Fig.

5C demonstrates $\zeta$ in terms of ITR contrast and heating time, and we found that for high ITR contrast ($R_2/R_1 > 25$), the temperature contrast is significantly enhanced with short heating time ($\zeta \sim 12\sim20$ when $Fo \sim 0.1$). Whereas for low ITR contrast ($R_2/R_1 < 25$), $\zeta$ is less sensitive to heating time, and for minimal ITR contrast ($R_2/R_1 \sim 1$-$2$), $\zeta$ approximately keeps constant ($\zeta \sim 1.5$) throughout the transient process. In conclusion, we demonstrate the combined affections of the ITRs, the geometries ($r_{JNP}$ and $\theta_0$), and the heating time on the temperature contrast. In general, smaller-sized JNP with high ITR contrast and moderate polar angle can be more sensitive to heating time and can lead to a greater temperature contrast with pulsed heating. As the first effort to shed light on the pulsed JNP heating during the transient process, we anticipate Fig. 4&5 to advance our understanding of the thermal transport with JNP heating, bringing inspiration for further work and providing route maps for developing novel applications.

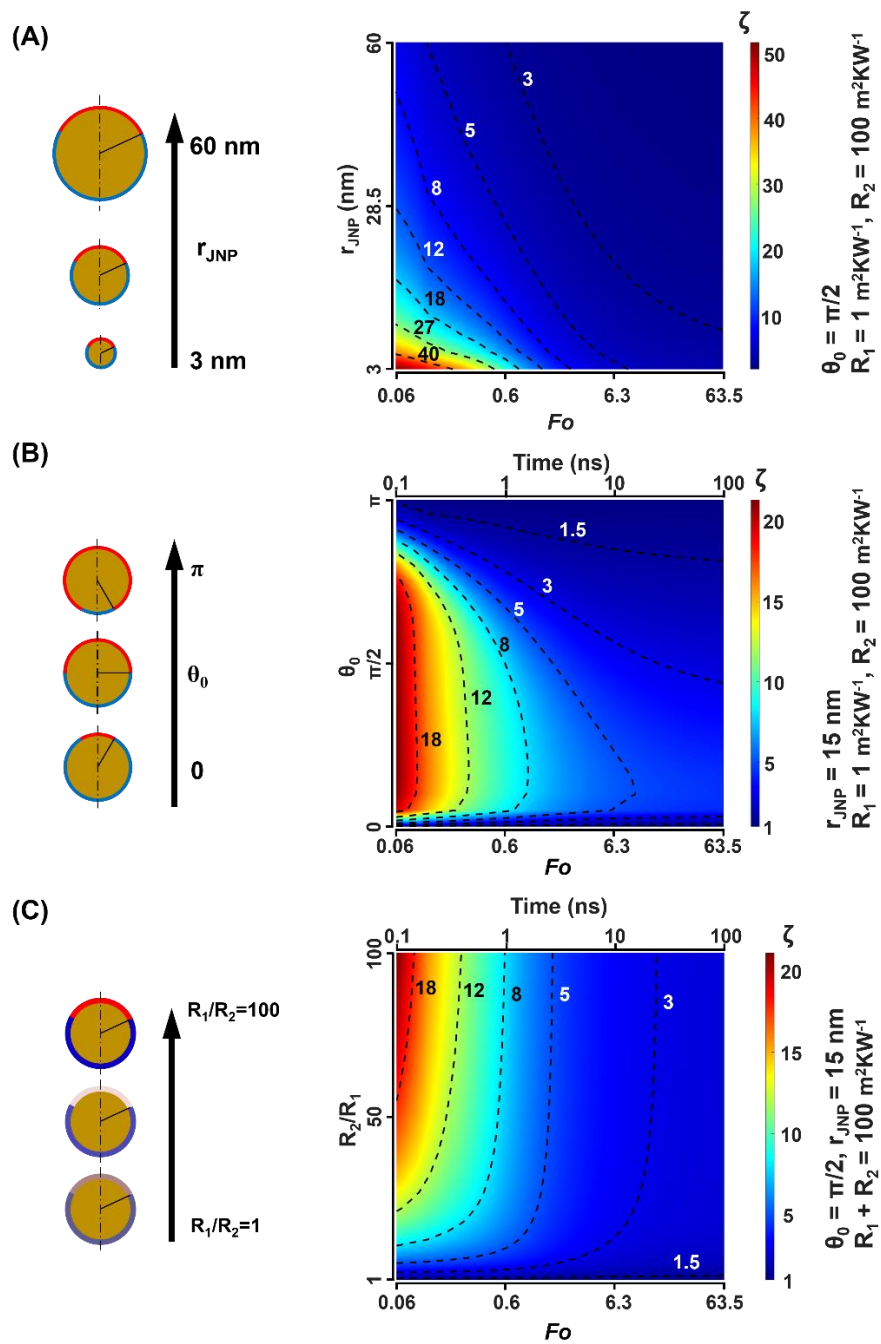

**Fig. 5 The effect of thermal resistance($R_2/R_1$), polar angle($\theta_0$), and size of the particle($r_{JNP}$) on temperature contrast during the transient process.** (A) $\zeta$ map in terms of ITR contrast ($R_2/R_1$) and heating time ($t$). Here we regulate $R_2 + R_1 = 100$ m$^2$KW$^-$

[1] and alter $R_2/R_1$ (B) $\zeta$ map in terms of $r_{JNP}$ and $Fo$. (C) $\zeta$ map in terms of $\theta_0$ and $t$. The dotted lines are contour lines for $\zeta$.

**Conclusion:**

In this work, we focused on thermal transport and directional heating with Janus nanoparticle (JNP) heating. We developed an FDM numerical framework and defined a dimensionless parameter ($\zeta \equiv \frac{\Delta T_1}{\Delta T_2}$) to characterize the temperature contrast. Based on this, we analyzed the effect of ITR contrast, $r_{JNP}$, and $\theta_0$ on temperature contrast under steady-state and transient heating. First, we found a significant size dependence of temperature contrast, where a smaller JNP can lead to greater temperature contrast. Additionally, we reveal the role of $\theta_0$ and demonstrate its asymmetric effect on temperature contrast, finding that the maximum $\zeta$ is observed at $\theta_0 \sim \pi/12$. Importantly, *we* shed light on the temperature evolution near a JNP during a transient heating process and found a significant enhancement of the temperature contrast with pulsed JNP heating. We expect our work can help us better understand thermal transport with JNP heating and advance the field with new inspirations.

**ACKNOWLEDGMENT**

I am grateful for the useful suggestions from Dr. Yaning Liu and Dr. Hui Ouyang. Research reported in this publication was supported by National Institute of General Medical Sciences and National Institute of Allergy and Infectious Diseases of the National Institutes of Health under award numbers R35GM133653 and R01AI151374, the National Science Foundation under award number 2123971, and the American Heart Association under award number 19CSLOI34770004. The content is solely the responsibility of the authors and does not necessarily represent the official views of funding agencies.

**ASSOCIATED CONTENT:**

Supporting information (Docx)

**AUTHOR INFORMATION:**

**Corresponding Author: Zhenpeng Qin\***

**Department of Mechanical Engineering, University of Texas at Dallas, 800 West Campbell Road EW31, Richardson, Texas 75080, United States.**

**E-mail: Zhenpeng.Qin@UTDallas.edu**

**ORICD: 0000-0003-3406-3045**

**Author Contributions**

C.X. developed the numerical model and did the computational work and analysis. B.W. and Z.Q. participated in conceptualization and analysis. All authors contributed to the writing of the manuscript. Z.Q. directed the research.

**FUNDING**

The research reported in this work was partially supported by the National Institute of General Medical Sciences (NIGMS) of the National Institutes of Health (award number R35GM133653), the Collaborative Sciences Award from the American Heart Association (award number 19CSLOI34770004), and the High-Impact/High-Risk Research Award from the Cancer Prevention and Research Institute of Texas (award number RP180846). The content is the sole responsibility of the authors and does not necessarily represent the official views of the funding agencies.

**NOMENCLATURE**

**Roman letters**

$R_1$      Interfacial thermal resistance along the north hemisphere of Janus nanoparticle, $m^2$ K $W^{-1}$

$R_2$      Interfacial thermal resistance along the south hemisphere of Janus nanoparticle, $m^2$ K $W^{-1}$

$k$      Thermal conductivity, W $m^{-2}$ $K^{-1}$

$k_{JNP}$      Thermal conductivity in Janus nanoparticle, W $m^{-2}$ $K^{-1}$

$k_w$      Thermal conductivity in water, W $m^{-2}$ $K^{-1}$

$T$      Temperature, K

$T_{JNP}$      Temperature in the Janus nanoparticle, K

| | | |
|---|---|---|
| $T_w$ | Temperature in water, K | |
| $g$ | Heating power, W m$^{-3}$ | |
| $c$ | Specific heat, J kg$^{-1}$ | |
| $t$ | Time, s | |
| $r$ | Radius, m | |
| $r_{1D}$ | Radius of the 1D zone, m | |
| $r_{JNP}$ | Radius of the Janus nanoparticle, m | |
| $r_{boundary}$ | Radius of the water domain, m | |
| $D$ | Thermal diffusivity, m² s$^{-1}$ | |
| $D_{JNP}$ | Thermal diffusivity in Janus nanoparticle, m² s$^{-1}$ | |
| $D_w$ | Thermal diffusivity in water, m² s$^{-1}$ | |
| $R$ | Interfacial thermal resistance, m²KW$^{-1}$ | |
| $R_1$ | Interfacial thermal resistance on the north hemisphere, m²KW$^{-1}$ | |
| $R_2$ | Interfacial thermal resistance on the southern hemisphere, m²KW$^{-1}$ | |
| $Ka$ | Kaptiza number, $Ka \equiv \dfrac{Rk}{r}$ | |
| $Ka_1$ | Kaptiza number on the north hemisphere, $Ka_1 \equiv \dfrac{R_1 k}{r_{JNP}}$ | |
| $Ka_2$ | Kaptiza number on the southerm hemisphere, $Ka_2 \equiv \dfrac{R_2 k}{r_{JNP}}$ | |
| $q$ | Heat flux, W | |

$q_1$     Heat flux through the north hemisphere, W

$q_2$     Heat flux through the southern hemisphere, W

$\dot{q}$     Heat flux rate, Wm$^{-2}$

$\dot{q}_1$     Heat flux rate at the north pole, Wm$^{-2}$

$\dot{q}_2$     Heat flux rate at the south pole, Wm$^{-2}$

$Fo$     Fourier number, $Fo \equiv \dfrac{Dt}{r_{JNP}^{2}}$

**Greek symbols**

$\theta$     Polar angle

$\theta_0$     The critical polar angle that pinpoints the boundary between the heterogeneous interfacial thermal resistances.

$\zeta$     Dimensionless parameter to characterize the temperature contrast, $\zeta \equiv \dfrac{\Delta T_1}{\Delta T_2}$

$\zeta_{s.s.}$     Dimensionless parameter to characterize the temperature contrast for steady-state.

$\rho$     Density, kg m$^{-3}$

$\tau$     Characteristic thermal relaxation time, $\tau \equiv \dfrac{r^2}{D}$, s

**Abbreviations**

JNP     Janus nanoparticle

ITR     Interfacial thermal resistance

MD     Molecular dynamic

FDM     Finite difference method

S.S.     Steady-state

# Supporting information for

**REGULATING NANOSCALE HEAT TRNSFER WITH JANUS NANOPARTICLES**

**This PDF file includes:**

Equations S1 to S3

Figures S1 to S9

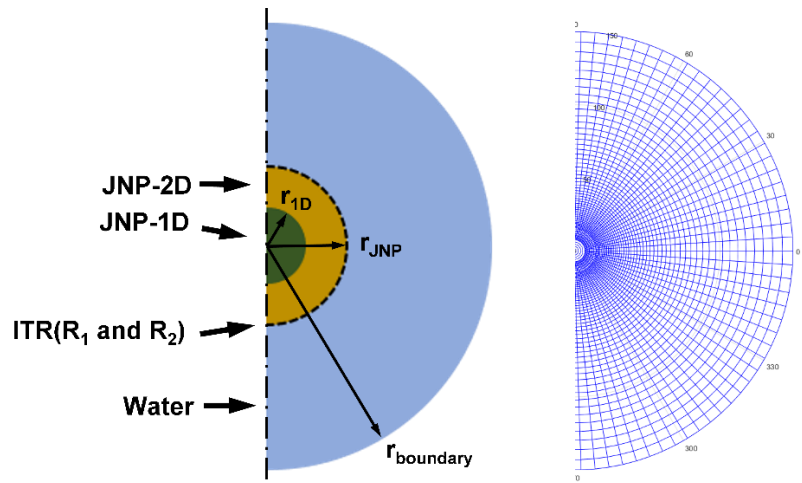

**Figure S1. The setup for the FDM model.** We expanded the governing equation in spherical coordination and discretized the system with an orthogonal mesh. To separate the singularities in the radius direction and the polar angle direction, we adopt a 1D zone with isothermal in the polar angle direction around the original point. Here we use $r_{1D}$ to denote the size of this 1D zone.

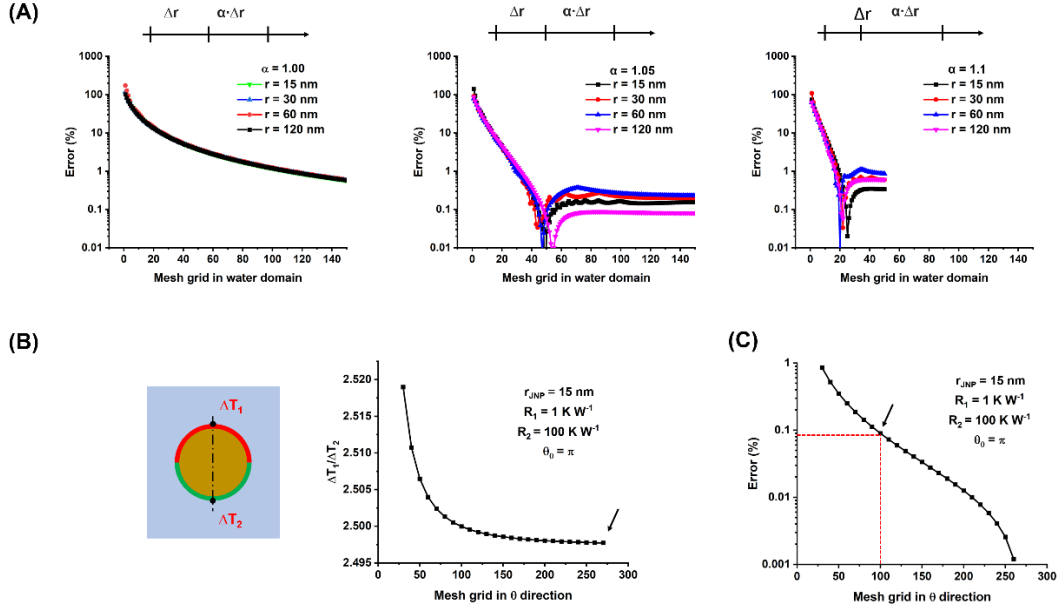

**Figure S2. The mesh independence analysis for radius direction.** (A) The truncation error in terms of mesh grid number in the radius direction under different biased mesh factor $\alpha$ ($\Delta r_{i+1} = \alpha \Delta r_i$). (B) Temperature contrast in terms of mesh grid number in the polar angle direction. (C) The truncation error in terms of f mesh grid number in the polar angle direction.

We validated our model by doing a comprehensive mesh dependency analysis. Here, we start with the mesh in the radius and the polar direction for the steady-state model. We assume uniform interfacial thermal resistance (ITR) along the JNP-water interface, thus making a homogenous NP heating where the analytical solution is available. As we have a linear biased mesh grid in the radius direction, we analyzed the truncation error by comparing the FDM results with the analytical solution with different biased mesh factor $\alpha$ ($\Delta r_{i+1} = \alpha \Delta r_i$) and mesh density. The error is calculated by the following equation:

$$error = \frac{abs(\Delta T_{analytical} - \Delta T_{FDM})}{\Delta T_{analytical}} \times 100\% \tag{S1}$$

As shown in Fig. S2A, when $\alpha = 1$, we have a uniform mesh, and the truncation error drops as we increase the mesh density; when the mesh grid number is 140, the error is around 1%. When $\alpha = 1.05$, we have a biased mesh, and the error drops much faster as the mesh density increases; when the mesh grid number is 50, the error is around 0.1%. This is because the biased mesh gives a higher mesh density near the JNP where the greatest temperature gradient is observed (Fig. S1), and thus greatly reduces the error and makes the mesh more efficient. As we continuously increase $\alpha$ ($\alpha = 1.1$), we found that the FDM results become less stable as we increase the mesh density. This might be because the over-biased mesh has coarse mesh grids near the boundary, which can bring extra error. In summary, we validate our model in the radius direction; we found a properly biased mesh factor ($\alpha = 1.05$) and a mesh density (50, minimum $\Delta r = 0.45$ nm); based on this setup, the truncation error is at the order of 0.1%.

Next, we did the mesh independence analysis in the polar angle ($\theta$) direction. Here we introduced a heterogeneous ITR and investigated the temperature contrast ($\frac{\Delta T_1}{\Delta T_2}$) with different mesh densities. Fig. S2B demonstrates that the value of $\frac{\Delta T_1}{\Delta T_2}$ converges as the mesh density increases. Here we assume the results at mesh grid = 275 is the accurate solution, and we calculate the error by comparing the results with this accurate solution (Fig. S2C); we could see the error is at the magnitude of 0.1% for the mesh grid = 100 ($\Delta\theta = \pi/100$).

We chose a truncation error of 0.1% that indicates sufficient accuracy for our mesh set-up; thus, we will keep the set-up ($\alpha$ = 1.05, minimum $\Delta r$ =0.45 nm, $\Delta\theta = \pi/100$) for the steady-state modeling throughout our paper.

Next, we analyzed the boundary effect. In reality, a free JNP can be treated as a single JNP surrounded by infinite water. In our model, we cannot have an infinite water domain, and the size of the boundary of the water domain can bring error, *i.e.*, boundary effect. Here we calculated the error of boundary effect by comparing the FDM result with a finite boundary radius and the analytical solution with an infinite boundary size. As shown in Fig. S3A, for boundary radius = 3000 nm (200 x $r_{JNP}$) the error of boundary effect drops to 1%, and we believe this is accurate for our analysis (Fig. S3B). We will keep this set-up throughout our paper.

Finally, we analyzed mesh independence and validated the model in the temporal direction. The spatial mesh set-ups are identical to the steady-state model. Here we plot the FDM results in terms of temporal mesh density (Fig. S4A), we found that the FDM results converge as mesh density increases, and we believe a temporal mesh interval of 0.1 ns can bring sufficiently accurate results. Here we further compared the FDM result with the analytical solution (Fig. S4B),[1] and demonstrated that the FDM results are consistent with the analytical solutions.

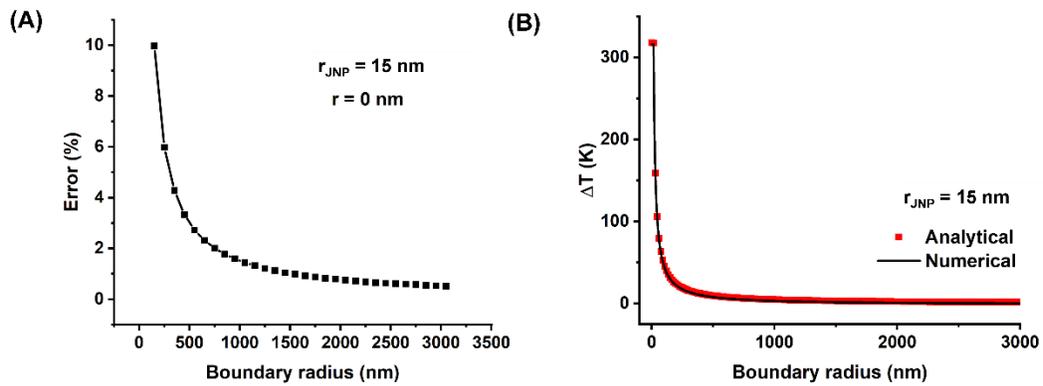

**Figure S3. The boundary effect analysis of the FDM model.** (A) The error of the temperature rise in JNP by comparing the result with an analytical model that has an infinite boundary size. (B) The temperature profile comparison between the analytical solution with infinite boundary size and FDM result with boundary size = 3000 nm.

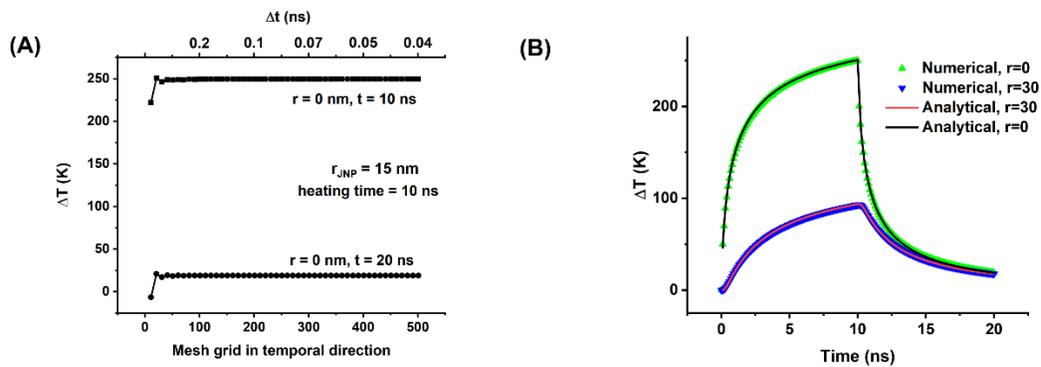

**Figure S4. The mesh independence for temporal direction.** (A) The mesh independence analysis in the temporal direction. (B) Comparison of FDM results with *Δt* = 0.1 ns and the analytical solution.

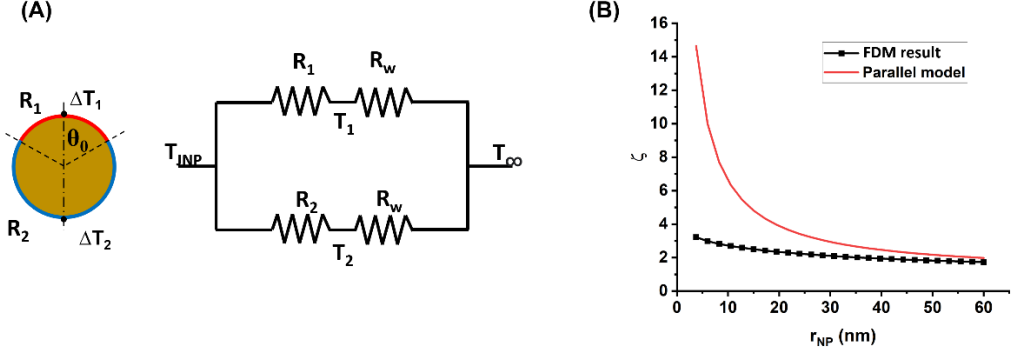

Figure S5. The parallel circuit model for the JNP heating under steady state.

For the parallel circuit model, the thermal transport with JNP can be described with a set of linear equations:

$$\begin{bmatrix} \dfrac{1}{R_1} & -\left(\dfrac{1}{R_1}+\dfrac{k_w}{r_{JNP}}\right) & 0 \\ \dfrac{1}{R_1} & 0 & -\left(\dfrac{1}{R_2}+\dfrac{k}{r_{JNP}}\right) \\ \left(\dfrac{A_1}{R_1}+\dfrac{A_2}{R_2}\right) & -\dfrac{A_1}{R_1} & -\dfrac{A_2}{R_2} \end{bmatrix} \begin{bmatrix} \Delta T_{JNP} \\ \Delta T_1 \\ \Delta T_2 \end{bmatrix} = \begin{bmatrix} Q \\ 0 \\ 0 \end{bmatrix}$$

$$A_1 = 2\pi r_{JNP}^2(1-cos(\theta_0))$$
$$A_2 = 2\pi r_{JNP}^2(1+cos(\theta_0))$$

(S2)

By solving Equation (S2), we have the temperature rise at the north pole and south pole:

$$\begin{bmatrix} \Delta T_{JNP} \\ \Delta T_1 \\ \Delta T_2 \end{bmatrix} = \begin{bmatrix} \dfrac{Q(r_{JNP}+R_1 k)(r_{JNP}+R_2 k)}{A_1 k r_{JNP}+A_2 k r_{JNP}+A_1 k^2 r_{JNP}+A_2 k^2 r_{JNP}} \\ \dfrac{Q r_{JNP}(r_{JNP}+R_2 k)}{A_1 k r_{JNP}+A_2 k r_{JNP}+A_1 k^2 r_{JNP}+A_2 k^2 r_{JNP}} \\ \dfrac{Q r_{JNP}(r_{JNP}+R_1 k)}{A_1 k r_{JNP}+A_2 k r_{JNP}+A_1 k^2 r_{JNP}+A_2 k^2 r_{JNP}} \end{bmatrix}$$

(S3)

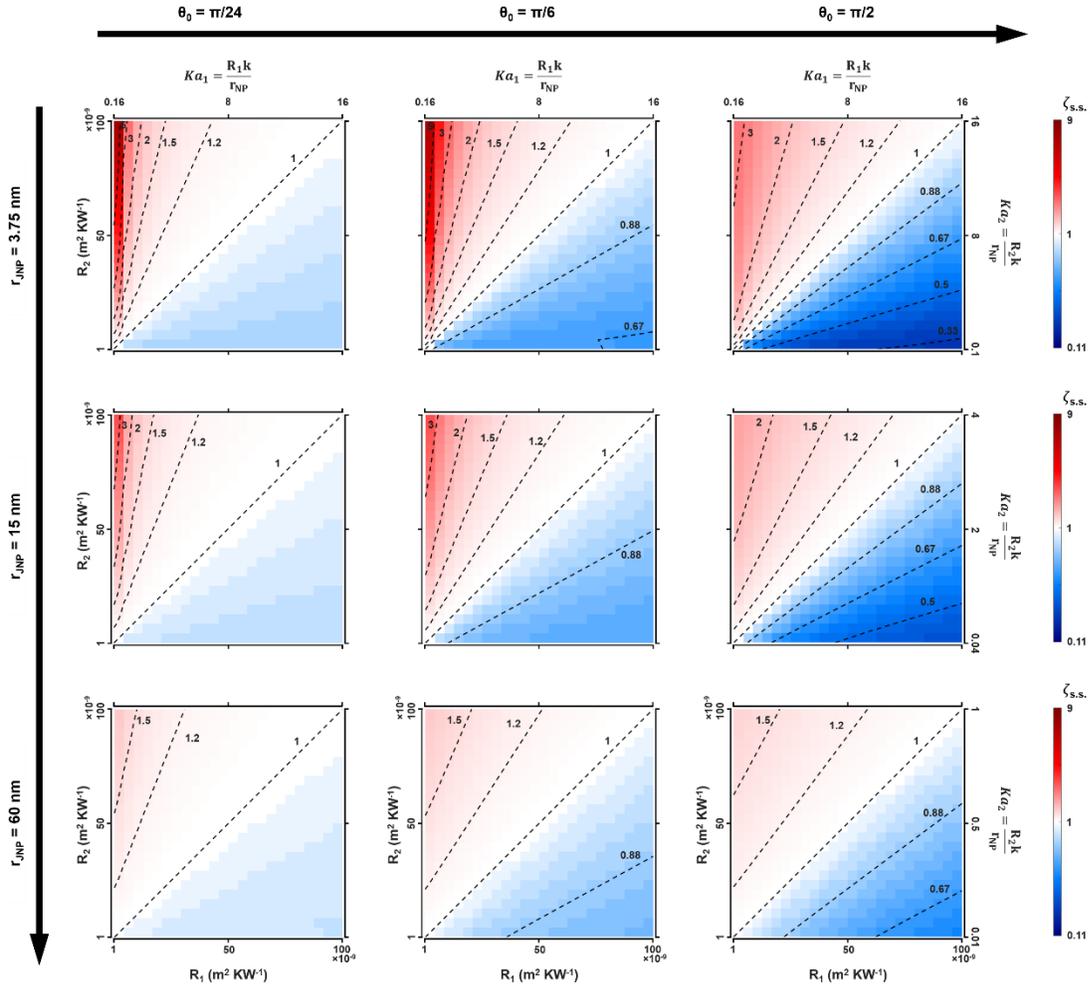

**Figure S6.** $\zeta_{s.s.}$ map in terms of $R_1$ and $R_2$ with different $r_{JNP}$ and $\vartheta_0$.

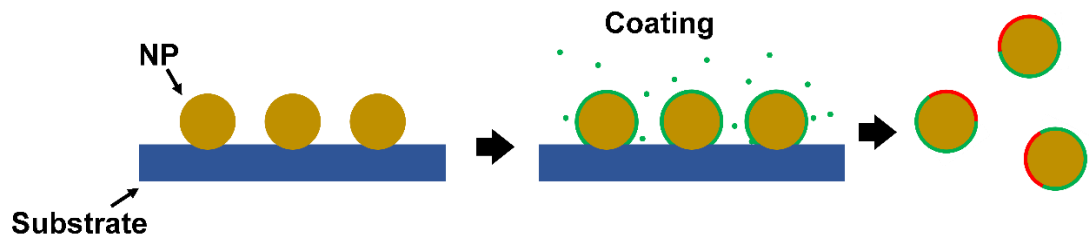

**Figure S7.** Schematic illustration of the synthesize protocol for JNP.

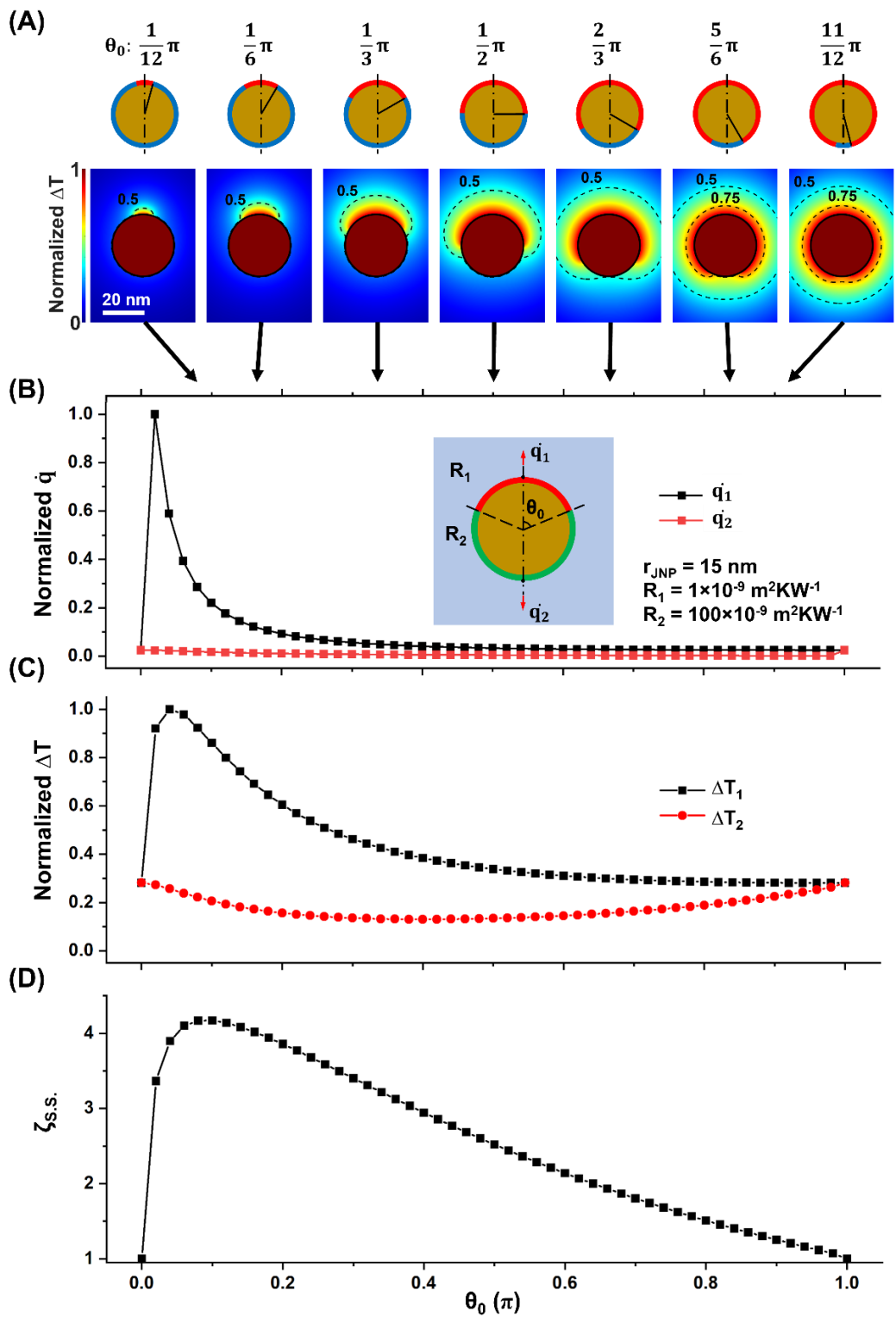

**Figure S8. The effect of the polar angle for heterogeneous coating ($\theta_0$) on the temperature profile.** (A) The $\Delta T$ profile of JNP heating with different $\theta_0$. $R_1 = 1 \times 10^{-9}$ m$^2$KW$^{-1}$ and $R_2 = 100 \times 10^{-9}$ m$^2$KW$^{-1}$, $r_{JNP} = 15$

nm. (B) The local heat flux rate across the interface at northern pole ($\dot{q}_1$) and southern pole ($\dot{q}_2$) in terms of $\theta_0$. (C) $\Delta T_1$ and $\Delta T_2$ in terms of $\theta_0$. (D) $\zeta_{s.s.}$ in terms of $\theta_0$.

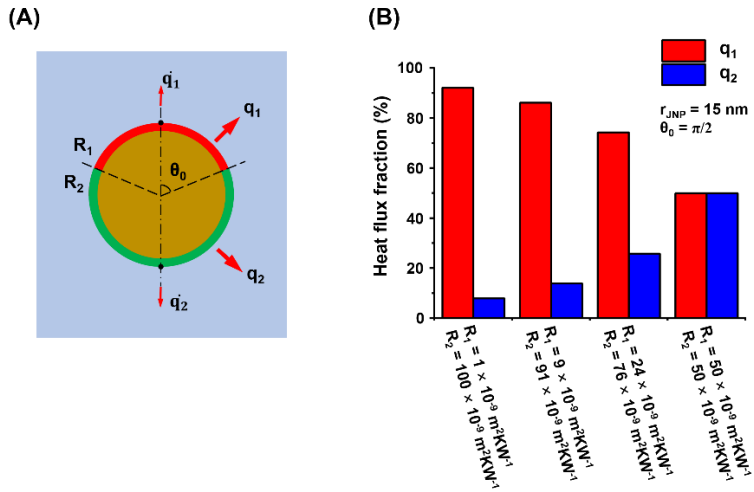

**Figure S9. The effect of ITR contrast on the heat flux fraction through the north and southern hemisphere.** (A) Schematic illustration of the heat flux through the north hemisphere ($q_1$) and south hemisphere ($q_2$). (B) The fraction of $q_1$ and $q_2$ with different ITR contrast.